\documentclass[pre,
preprint,
dvips,
aps]{revtex4}
\usepackage{amsmath,amssymb,amsfonts,graphicx}

\begin{document}

\title{Nonequilibrium Stefan-Boltzmann law\thanks{%
the contents of this paper has been presented at JETC} }
\author{Agustin P\'{e}rez-Madrid}
\email{agustiperezmadrid@ub.edu}
\affiliation{Departament de F\'{\i}sica Fonamental, Facultat de F\'{\i}sica, Universitat
de Barcelona, Av. Diagonal 647, 08028 Barcelona, Spain}
\author{J.~Miguel Rub\'{\i}}
\affiliation{Departament de F\'{\i}sica Fonamental, Facultat de F\'{\i}sica, Universitat
de Barcelona, Av. Diagonal 647, 08028 Barcelona, Spain}
\author{Luciano C. Lapas}
\affiliation{Instituto de F\'{\i}sica and Centro Internacional de F\'{\i}sica da Mat\'{e}%
ria Condensada, Universidade de Bras\'{\i}lia, Caixa Postal 04513, 70919-970
Bras\'{\i}lia, Distrito Federal, Brazil}

\begin{abstract}
We study thermal radiation outside equilibrium. The situation we consider
consists of two bodies emitting photons at two different temperatures. We
show that the system evolves to a stationary state characterized by an
energy current which satisfies a law similiar to the Stefan-Boltzmann law.
The magnitude of this current depends on the temperature of the emitters
expressed through the difference of the fourth power of these
tempereatures.The results obtained show how the classical laws governing the
thermal radiation at equlibrium can be generalized away from equilibrium
situations.
\end{abstract}

\maketitle

\section{Introduction}

Thermal radiation at equilibrium was studied by Planck \cite{Planck} by
using equilibrium thermodynamic concepts. The thermal properties of the gas
of photons are well-known. One of them, the Stefan-Boltzmann law \cite%
{pathria} gives the value of the energy flux in terms of the temperature of
the emitter through a power law $\sigma T^4$.

However, in many instances frequently found in nano-systems, the radiation
is not in equilibrium due to the presence of thermal sources or temperature
gradients. This is what happens in nanostructures such as solar cells and
thermophotovoltaic devices \cite{zhang,vos}, even in cancer therapies \cite%
{rabin,fortin}, just to cite some exemples. In these situations, the
classical scheme is no longer applicable and it then becomes necessary to
employ a nonequilibrium theory. A first attempt to describe non-equilibrium
radiation could be performed via nonequilibrium thermodynamics \cite{deGroot}%
. Nevertheless, some of the laws governing the behavior of thermal radiation
are non-linear laws whose derivation is beyond the scope of this theory
which provides only linear relationships between fluxes and forces.

In this article, we will show that this limitation can be overcome if
mesoscopic nonequilibrium thermodynamics \cite{reguera,rubi} is used. This
can be done by performing a description in terms of an internal variable 
\cite{agusti,rubi2} --the momenta of the photons-- and by assuming local
equilibrium in phase--space. In this way it is possible to analyze the
underlying activated process for which photons are emitted. The photon
energy current does obtained leads to the nonequilibrium
Stefan-Boltzmann law.

This article is organized as follows. In Section 2, we study the photon gas
outside equilibrium. Starting from the Gibbs entropy postulate, we derive
the entropy production and from this the current of photons. In Section 3,
we analyze the steady state obtained when photons are emitted by two bodies
at two different temperatures. By using mesoscopic nonequilibrium
thermodynamics we derive the equivalent to the Stefan-Boltzmann law in this
nonequilibrium situation. In the conclusion, we discuss some perspectives of
the results obtained.

\section{The nonequilibrium photon gas}

Let us consider a gas of photons distributed through the law $n(\mathbf{%
\Gamma ,}t)$ which is the probability density defined in the single-particle
phase-space $\mathbf{\Gamma }=(\mathbf{p}$,$\mathbf{x})$, where $\mathbf{p}$%
, $\mathbf{x}$ are the momentum and position of a photon, respectively.
According to the principle of the conservation of probability, we assume
that $n(\mathbf{\Gamma ,}t)$ satisfy the continuity equation%
\begin{equation}
\frac{\partial }{\partial t}n(\mathbf{\Gamma ,}t)=-\frac{\partial }{\partial 
\mathbf{\Gamma }}\cdot \mathbf{J}\left( \mathbf{\Gamma ,}t\right)
\label{continuity}
\end{equation}%
where $\partial /\partial \mathbf{\Gamma }=(\partial /\partial p,\nabla )$, $%
\nabla =\partial /\partial x$. The continuity equation (\ref{continuity})
defines the probability current $\mathbf{J}\left( \mathbf{\Gamma ,}t\right)
=(J_{x},J_{p})$ which must be determined by means of the methods of
nonequilibrium thermodynamics.

According to the Gibbs entropy postulate, our thermodynamic analysis of the
nonequilibrium gas is based in the assumption of the density functional 
\begin{equation}
S(t)=-k_{B}\int n(\mathbf{\Gamma ,}t)\ln \frac{n(\mathbf{\Gamma ,}t)}{n_{eq}(%
\mathbf{\Gamma })}d\mathbf{\Gamma +}S_{0}  \label{entropy}
\end{equation}%
as the nonequilibrium entropy of the system \cite{deGroot,reguera,rubi}.
Here, $S_{0}$ is the equilibrium entropy of the gas plus the thermal bath
and $n_{eq}(\mathbf{\Gamma })$ is the equilibrium probability density
function.

By taking variations of Eq.\ (\ref{entropy}) we obtain

\begin{equation}
\delta S=-k_{B}\int \delta n(\mathbf{\Gamma ,}t)\ln \frac{n(\mathbf{\Gamma ,}%
t)}{n_{eq}(\mathbf{\Gamma })}d\mathbf{\Gamma }\text{ \ ,}  \label{gibbs_eq}
\end{equation}%
where once one introduces the nonequilibrium chemical potential 
\begin{equation}
\mu (\mathbf{\Gamma ,}t)=k_{B}T\ln \frac{n(\mathbf{\Gamma ,}t)}{n_{eq}(%
\mathbf{\Gamma })}\text{, }  \label{chem_pot}
\end{equation}%
Eq.\ (\ref{gibbs_eq}) can be written 
\begin{equation}
\delta S=-\int \frac{\mu (\mathbf{\Gamma ,}t)}{T}\delta n(\mathbf{\Gamma ,}%
t)d\mathbf{\Gamma }\text{ \ .}  \label{variation}
\end{equation}%
Here, Eq. (\ref{variation}) which is the Gibb's equation of thermodynamics
formulated in the phase-space, illustrates the physical meaning of
the nonequilibrium chemical potential (\ref{chem_pot}). Since $-T\delta
S=\delta F$, with $F$ the nonequilibrium free energy, Eq. (\ref{variation}) leads to

\begin{equation}
\delta F=\int \mu (\mathbf{\Gamma ,}t)\delta n(\mathbf{\Gamma ,}%
t)d\mathbf{\Gamma }\text{ \ .}  \label{variation2}
\end{equation}
On the other hand, the equilibrium chemical potential of photons is zero, thus no reference value $\mu_0$ is needed in Eq. (\ref{chem_pot}). From Eqs. (\ref{continuity}) and (\ref{variation}) we obtain the entropy
production%
\begin{equation}
\frac{\partial S}{\partial t}=-\int \mathbf{J}\left( \mathbf{\Gamma ,}%
t\right) \cdot \frac{\partial }{\partial \mathbf{\Gamma }}\frac{\mu (\mathbf{%
\Gamma ,}t)}{T}d\mathbf{\Gamma }\text{ \ ,}  \label{entropy_prod}
\end{equation}%
as the product of a thermodynamics current and the conjugated thermodynamic
force $\partial /\partial \mathbf{\Gamma (}\mu (\mathbf{\Gamma ,}t)/T)$ \cite%
{deGroot}. As usual in nonequilibrium thermodynamic \cite{deGroot}, from Eq.
(\ref{entropy_prod}) we derive the phenomenological law%
\begin{equation}
\mathbf{J}\left( \mathbf{\Gamma ,}t\right) =-\mathbf{L}(\mathbf{\Gamma }%
)\cdot \frac{\partial }{\partial \mathbf{\Gamma }}\frac{\mu (\mathbf{\Gamma ,%
}t)}{T}\text{ \ ,}  \label{pheno_1}
\end{equation}%
where $\mathbf{L}(\mathbf{\Gamma })$ is the matrix of phenomenological
coefficients. Hence, Eq. (\ref{pheno_1}) enables us to write the entropy
production Eq. (\ref{entropy_prod}) as a bilinear form%
\begin{equation}
\frac{\partial S}{\partial t}=\int \frac{\partial }{\partial \mathbf{\Gamma }%
}\frac{\mu (\mathbf{\Gamma ,}t)}{T}\cdot \mathbf{L}(\mathbf{\Gamma })\cdot 
\frac{\partial }{\partial \mathbf{\Gamma }}\frac{\mu (\mathbf{\Gamma ,}t)}{T}%
d\mathbf{\Gamma }\text{ .}  \label{entropy_prod_2}
\end{equation}%
In terms of the diffusion matrix $\mathbf{D}(\mathbf{\Gamma })=k_{B}\mathbf{M%
}$, where $\mathbf{M}(\mathbf{\Gamma })=\mathbf{L}/n$ is the mobility, Eq. (%
\ref{pheno_1}) can be written in a more convenient form%
\begin{equation}
\mathbf{J}\left( \mathbf{\Gamma ,}t\right) =-\mathbf{D}(\mathbf{\Gamma }%
)\cdot \frac{\partial }{\partial \mathbf{\Gamma }}n(\mathbf{\Gamma ,}t)\text{%
,}  \label{pheno_2}
\end{equation}%
which constitutes the Fick's law of diffusion \cite{landau} formulated in
the single-particle phase-space.

\section{Stationary state and the nonequilibrium Stefan-Boltzmann law}

In this section we will study the heat exchange by thermal radiation between
two bodies at different temperatures. This process comes into play in
nanostructures such as solar cells and thermophotovoltaic devices, just to
cite some exemples. Applications of thermophotovoltaic devices range from
hybrid electric vehicles to power sources for microelectronic systems \cite%
{zhang}. Therefore, as mentioned, the recent developments of nanotechnology
made the study of heat exchange by thermal radiation an object of growing
interest \cite{abramson}.

Hence, to undertake this study, we assume that the dynamics of the photons
is the result of two simultaneous processes: emission and absortion of cold
photons at $T_{C}$ and emission and absorption of hot photons at $T_{H}$,
this is illustrated by the Figure (see Fig. (1))

\begin{figure}[h]
\includegraphics[scale=1.3]{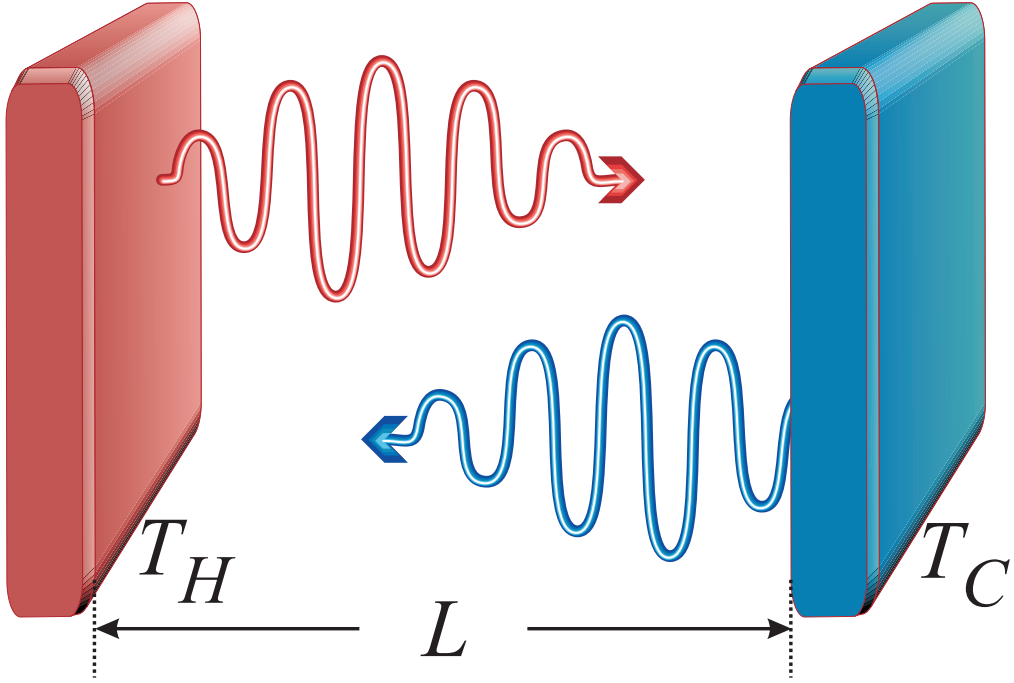}
\caption{(Color online) Schematic illustration of the radiation exchanged
between two materials maintained at different temperatures, $T_{H}$ and $%
T_{C}$, separated by a distance $L$.}
\label{fig1}
\end{figure}

Since the photons do not interact among themselves we assume that the system
is homogeneous, $\mathbf{\Gamma \longrightarrow p}$, \ and the diffusion
matrix reduces to a scalar $D(\mathbf{\Gamma })$, the diffusion coefficient.
Additionally, if there are only hot and cold photons 
\begin{equation}
\mathbf{J}\left( \mathbf{p,}t\right) =\widehat{\mathbf{J}}_{C}(t)\delta (%
\mathbf{p}-\mathbf{p}_{C})+\widehat{\mathbf{J}}_{H}(t)\delta (\mathbf{p}-%
\mathbf{p}_{H})\text{ ,}  \label{current_pre}
\end{equation}%
\textit{i.e. }the system reaches a state of quasi equilibrium. Thus,
integration of Eq. (\ref{pheno_2}) taking into account Eq. (\ref{current_pre}%
) leads to%
\begin{equation}
\frac{\widehat{J}_{C}(t)}{D_{C}}+\frac{\widehat{J}_{H}(t)}{D_{H}}=-\left[ n(%
\mathbf{p}_{C}\mathbf{,}t)-n(\mathbf{p}_{H}\mathbf{,}t)\right] \text{ \ ,}
\label{current}
\end{equation}%
with $J(t)=\mathbf{u}\cdot \mathbf{J}(t)$ and $\mathbf{u}$ is the unit
vector normal to the walls. Whence, introducing the net current $J(t)$
defined through%
\begin{equation}
\frac{J(t)}{aD_{C}D_{H}}=\frac{\widehat{J}_{C}(t)}{D_{C}}+\frac{\widehat{J}%
_{H}(t)}{D_{H}}\text{ \ ,}  \label{current_3}
\end{equation}%
or equivalently 
\begin{equation}
J(t)=aD_{H}\widehat{J}_{C}(t)+aD_{C}\widehat{J}_{H}(t)\text{ ,}
\label{current_6}
\end{equation}%
we can rewrite Eq. (\ref{current})%
\begin{equation}
J(t)=-aD_{C}D_{H}\left[ n(\mathbf{p}_{C}\mathbf{,}t)-n(\mathbf{p}_{H}\mathbf{%
,}t)\right] \text{ ,}  \label{current_4}
\end{equation}%
\newline
where $a$ is an effective parameter accounting for the dimensionality. Here, term by term comparison of \ Eqs. (\ref{current_6}) and (\ref%
{current_4}) leads to the identification 
\begin{align}
\widehat{J}_{C}(t)& =-aD_{C}n(\mathbf{p}_{C}\mathbf{,}t),  \label{flow_1} \\
\widehat{J}_{H}(t)& =aD_{H}n(\mathbf{p}_{H}\mathbf{,}t)\text{ \ .}
\label{flow_2}
\end{align}%
Therefore, 
\begin{equation}
-D_{H}\widehat{J}_{C}(t)=D_{H}D_{C}n(\mathbf{p}_{C}\mathbf{,}t)  \label{one}
\end{equation}%
represents the fraction of photons absorved at the hot wall from the
fraction $J_{C}(t)$ of \ photons emited at the cold wall. In the same way, 
\begin{equation}
D_{C}\widehat{J}_{H}(t)=D_{C}D_{H}n(\mathbf{p}_{H}\mathbf{,}t)  \label{two}
\end{equation}%
represents the fraction of photons absorved at the cold wall from the
fraction $J_{H}(t)$ of photons emited at the hot wall. \newline
To better illustrate the physics under the nonequilibrium process involving
the photons, we introduce the affinity%
\begin{equation}
A=\mu _{C}-\mu _{H},  \label{affinity}
\end{equation}%
where $\mu _{C}=$ $\mu (\mathbf{p}_{C}\mathbf{,}t)$ and $\mu _{H}=\mu (%
\mathbf{p}_{H}\mathbf{,}t)$, which with the help of Eq. (\ref{chem_pot}),
enables us to rewrite Eq. (\ref{current_4}) as \cite{agusti} 
\begin{equation}
J(t)=aD_{C}D_{H}\exp \left( \frac{\mu _{H}}{k_{B}T_{H}}\right) \left( 1-\exp
\left( \frac{A}{k_{B}T_{H}}\right) \exp \left[ \frac{\mu _{H}}{k_{B}}\left( 
\frac{1}{T_{C}}-\frac{1}{T_{H}}\right) \right] \right) .  \label{current_7}
\end{equation}%
This result shows the nonlinearity inherent to the process involving the radiation since this corresponds to a heat current which does not satisfies the Fourier law. 

In general, the diffusion coefficient might depend on the frequency.
However, one can introduce a cut-off \ limit, $\lambda _{T}=c\hslash /k_{B}T$%
, the thermal wavelength of a photon, which marks this dependence. So, for
length scales $L\gg \lambda _{T}$ (\textit{i.e. }high frequencies) we can
treat the photons as point particles, hence in this case we deal with the
free diffusion of point particles for which the diffusion coefficient is
constant. On the other hand, when $L\lesssim \lambda _{T}$ (\textit{i.e. }%
low frequencies) we deal with the problem of cage-diffusion \cite{pusey},
therefore the diffusion coefficient must depend on the ratio $\lambda _{T}/L$
or equivalently on the frequency \cite{agusti3,greffet}.

In the stationary state, the current of photons per unit of volume in phase-space is given by %
\begin{equation}
J_{st}(\omega )=aD_{C}(\omega )D_{H}(\omega )\left[ n(\omega
,T_{H})-n(\omega ,T_{C})\right] \text{ ,}  \label{current_2}
\end{equation}%
where%
\begin{equation}
n(\omega ,T)=2\frac{N(\omega ,T)}{h^{3}},  \label{current_21}
\end{equation}%
with $h$ being the Planck constant and $N(\omega ,T)$ the averaged number of
particles in a elementary cell of the phase-space given by the Planck
distribution \cite{pathria} 
\begin{equation}
N(\omega ,T)=\frac{1}{\exp \left( \hslash \omega /kT\right) -1}.
\label{planck}
\end{equation}%
In Eq. (\ref{current_21}) the factor 2 comes from the polarization of the
photons. Here, the diffusion constant plays the role of the effective cross
section.

By multiplying Eq. (\ref{current_2}) by $\hslash \omega $ --the energy of a
photon-- and integrating in momentum space we obtain the heat current 
\begin{equation}
Q=\int \hslash \omega J_{st}(\omega )d\mathbf{p},  \label{heat_curr}
\end{equation}%
where $\mathbf{p}=\left( \hslash \omega /c\right) \mathbf{\Omega }_{p}$. In
the case $L\gg \lambda _{T}$ , and in an ideal situation, \ we can take $%
D_{C}=D_{H}=D=1$. Thus,  with Eq. (\ref{current_21}), Eq. (\ref{heat_curr})
leads to 
\begin{equation}
Q=\frac{\hslash }{4\pi ^{3}c^{3}}\int d\omega d\mathbf{\Omega }_{p}\omega
^{3}J_{st}(\omega )=\sigma \left( T_{H}^{4}-T_{C}^{4}\right) \text{ ,}
\label{integra}
\end{equation}%
where we have taken $a=c/4$, with $c$ the velocity of the radiation and $\sigma =\pi ^{2}k_{B}^{4}/60\hslash
^{3}c^{2}$ being the Stefan constantn \cite{pathria}. At equilibrium $T_{H}=T_{C}$, therefore $Q=0$.

For a s-dimensional system, the phase-space is made up of s coordinates and
s conjugated momenta. Hence, the volume of an elementary cell of this
phase-space is $h^{s}$ and now

\begin{equation}
n(\omega ,T)=2\frac{N(\omega ,T)}{h^{s}}.  \label{A1}
\end{equation}%
In this case, by integrating over a hypersphere in momentum-space 
and assuming that now $a$ is a function of $s$, $a(s)$, it this possible to
obtain%
\begin{equation}
Q=\sigma (s)\left( T_{H}^{s+1}-T_{C}^{s+1}\right)  \label{A2}
\end{equation}%
which generalizes results previously derived in Ref. \cite{landsberg}

\section{Conclusions}

In this paper we have performed a thermodynamic analysis of the radiative
heat exchange between two bodies at different temperatures separated by a
certain distance $L$. We have shown how the Stefan-Bolztmann law can be
generalized to nonequilibrium situations as those in which thermal radiation
is composed by photons emitted at two different temperatures and thus having
different momenta. In the framework of mesoscopic nonequilibrium
thermodynamics based on the Gibb's entropy postulate, we have obtained the
Gibb's equation of thermodynamics which here, describes the local
equilibrium in the phase-space. Then, by means of the usual procedure of
nonequilibrium thermodynamics, from the Gibb's equation we have derived the
Fick's law for the diffusion of photons valid at short time scales.

The exchange of energy is assumed to be due to activated processes related
to reaction rate currents which are derivable in the framework of our
thermodynamic theory. These currents provide us with the rate of absortion
and emission of photons. Since in the stationary state both hot and cold
photons are in local equilibrium, their rates are proportional to the Planck
distribution. Finally, the net current of heat is given through a balance of
these rate currents after integration over frequencies, constituting the
nonequilibrium Stefan-Boltzmann law.

Systems outside equilibrium exhibit peculiar features not observed in an
equilibrium state \cite{dufty}. The results obtained show how
non-equilibirum phenomena taking place in gases composed of quasi-particles,
governed by non-linear laws, can be analyzed by means of the methods of
nonequilibrium thermodynamics applied to the mesoscale. These methods have
been shown to be very useful in the study of activated processes of
different natures \cite{agusti2,lapas2}.

\subsection*{Acknowledgments}

This work was supported by the DGiCYT of Spanish Government under Grant No.
FIS2008-04386, and by Brazilian Research Foundation: CNPQ.


\begin{thebibliography}{99}
\bibitem{Planck} Planck, M., Treatise on Thermodynamics, Dover, New York,
1945.

\bibitem{pathria} Pathria, R.K., Statistical mechanics,\textit{\ }Pergamon
Press, Oxford,1988.

\bibitem{zhang} Zhang, Z., Nano/Microscale Heat Transfer, Nanoscience and
Technology Series, McGraw-Hill, New York, 2007.

\bibitem{vos} De Vos, A., Thermodynamics of Solar Energy Conversion,
WILEY--VCH Verlag, 2008.

\bibitem{rabin} Rabin, Y., Nanotechnology Essays, EurekAlert--Nanotechnology
In Context, August 2002.

\bibitem{fortin} Fortin, J.-P., Gazeau, F., Wilhelm, C., Intracellular
heating of living cells through N\'{e}el relaxation of magnetic
nanoparticles, Eur. Biophys. J., \textbf{37} (2008) 223--228.

\bibitem{dufty} Dufty, J. W., Rub\'{\i}, J. M., Generalized Onsager
symmetry, Phys. Rev. A, \textbf{36} (1987) 222--225 .

\bibitem{deGroot} ~de~Groot, S.~R.,~Mazur, P. Non-Equilibrium
Thermodynamics.\ Dover, New York, 1984.

\bibitem{reguera} Reguera, D., Vilar, J.M.G., Rub\'{\i}, J.M., The
Mesoscopic Dynamics of Thermodynamic Systems, J. Phys. Chem. B \textbf{109}
(2006), 21502--21515.

\bibitem{rubi} Vilar, J.M.G., Rub\'{\i}, J.M., Thermodynamics "beyond" local
equilibrium, Proc. Natl. Acad. Sci., \textbf{98} (2001), 11081--11084.

\bibitem{agusti} Pagonabarraga, I., P\'{e}rez-Madrid, A., Rub\'{\i}, J.M.,
Fluctuating hydrodynamics approach to chemical reactions, Physica A, \textbf{%
237} (1997), 205--219.

\bibitem{rubi2} Rub\'{\i}, J.M., P\'{e}rez-Madrid, A., Physica A, \textbf{264%
} (1999), 492--502.

\bibitem{vanKampen} ~van~ Kampen, N.~G., Stochastic Processes in Physics and
Chemistry, \ North-Holland, Amsterdam, 1990.

\bibitem{landau} ~Landau, L.~D., ~Lifshitz, E.~M., Statistical Physics,
Course of Theoretical Physics, Vol. 9, Part 2, Pergamon Press, Oxford, 1980.

\bibitem{abramson} Abramson, A.R., Tien, C.L., Recent developments in
microscale thermophysical engineering, Nanosacle Microscale Thermophys.
Eng., \textbf{3} (1999) 229--244.

\bibitem{pusey} Pusey, P.N., Lekkerkerker, H.N.W., Cohen, E.G.D., de
Schepper, I.M., Analogies between the dynamics of concentrated charged
colloidal suspensions and dense atomic liquids Physica A, \textbf{164}
(1990), 12--27.

\bibitem{agusti3} P\'{e}rez-Madrid, A., Rub\'{\i}, J. M., Lapas, L.C., Heat
transfer between nanoparticles: Thermal conductance for near-field
interactions, Phys. Rev. B, \textbf{77} (2008), 155417(1)--155417(7).

\bibitem{greffet} Domingues, G., Volz, S., Joulain, K., Greffet, J.J., Heat
Transfer between Two Nanoparticles Through Near Field Interaction, Phys.
Rev. Lett., \textbf{94} (2005), 085901(1)--085901(4).

\bibitem{agusti2} P\'{e}rez-Madrid, A., Reguera, D., Rub\'{\i}, J. M.,
Origin of the violation of the fluctuation--dissipation theorem in systems
with activated dynamics , Physica A, \textbf{329} (2003), 357--364.

\bibitem{lapas2} A. P\'{e}rez-Madrid, L.C. Lapas, J. M. Rub\'{\i}, Heat
Exchange between Two Interacting Nanoparticles beyond the
Fluctuation-Dissipation Regime, Phys. Rev. Lett., \textbf{103} (2009),
048301(1)--048301(4).

\bibitem{landsberg} Landsberg, P.T., De Vos, A., The Stefan-Boltzmann
constant in n-dimensional space, J. Phys. A: Math. Gen., \textbf{22} (1989),
1073--1084.
\end{thebibliography}
\end{document}